\documentclass[final]{aipproc}

\usepackage{amsmath}
\layoutstyle{6x9}
\begin{document}

\title{Renormalization group evolution\\ of the CKM matrix\thanks{
Talk given in the X Mexican School of Particles and Fields,
Playa del Carmen, M\'exico, 2002}}

\author{P.~Kielanowski}
{address={Departamento de F\'{\i}sica, Centro de Investigaci\'{o}n y
de Estudios Avanzados del IPN,\\ 07360 M\'{e}xico D.F., Mexico}
}
\author{S.R.~Ju\'{a}rez~W.}
{address={Departamento de F\'{\i}sica, Centro de Investigaci\'{o}n y
de Estudios Avanzados del IPN,\\ 07360 M\'{e}xico D.F., Mexico},
altaddress={Departamento de F\'{\i}sica, Escuela Superior de
F\'{\i}sica y Matem\'{a}ticas,\\
Instituto Polit\'{e}cnico Nacional, 07738 M\'{e}xico D.F., Mexico}
}

\begin{abstract}
We present here the most important ideas, equations and solutions for the
running of all the quark Yukawa couplings and all the elements of the
Cabibbo-Kobayashi-Maskawa matrix, in the approximation of one loop, and up
to order $\lambda ^{4}$, where $\lambda \sim 0.22$ is the sine of the
Cabibbo angle. Our purpose is to determine what the evolution of these
parameters may indicate for the physics of the standard model (SM), the
minimal supersymmetric standard model (MSSM) and for the Double Higgs Model
(DHM).

\end{abstract}

\maketitle


The number of free parameters of the standard model is at least 18. For a
fundamental theory this is a rather large number of parameters. 11 of these
parameters are related to the masses, 3 are the coupling constants and 4 are
the parameters of the Cabibbo-Kobayashi-Maskawa (CKM) matrix. All these
parameters have to be given in the Lagrangian of the standard model which is
also a function of the fields. We are accustomed to talk about the masses
and coupling constants but the parameters of the Lagrangian are not really
the physical observables that are measured in the experiment. The real
physical observables are the matrix elements of the $S$ matrix which are
rather complicated functions of the Lagrangian's parameters. The physical
particle masses are for example determined from the positions of the poles
of the $S$ matrix elements. Each physical theory requires the renormalization
in order to become finite. The renormalization is the procedure that links
the physical observables with the parameters of the Lagrangian. In realistic
theories the renormalization also serves as a method to remove the
infinities that appear in the loop diagrams. The renormalization procedure
first employs the regularization and then by using the appropriate counter
terms, the infinities are removed from the observables. In the process of
the renormalization one has to impose some conditions that introduce into
the theory an arbitrary constant $\mu$ of the dimension of energy which is
called the renormalization scale. The observables cannot and do not depend
on the renormalization scale but the coupling constants and masses in the
Lagrangian do. Their dependence on the renormalization scale $\mu $ is
determined from the renormalization group 
equations~\cite{ref1,ref2,ref3,ref4,ref7,ref5,n2,n3,n4,n6,n7}
 that are derived from
the condition of independence of the observables on the renormalization
scale.

The equations of the renormalization group give the dependence of the
parameters of the Lagrangian on the renormalization scale. The first
interesting result is that the coupling constants of QED and QCD have the
following dependence on the renormalization scale 
\begin{eqnarray}
\text{QED} &:&\text{{}}\;\;\;\;\mu \nearrow \;\;\longrightarrow \alpha _{%
\text{QED}}\nearrow  \label{a1} \\
\text{QCD} &:&\text{{}}\;\;\;\;\mu \nearrow \;\;\longrightarrow \alpha _{%
\text{QCD}}\searrow
\end{eqnarray}
The decreasing of the QCD coupling constant as the renormalization scale $%
\mu $ increases is called asymptotic freedom and historically was one of the
most important arguments for the introduction of the QCD, but we also see
from the previous equation that there must exist a scale $\mu $ where the
two coupling constants are equal 
\begin{equation}
\alpha _{\text{QED}}=\alpha _{\text{QCD}}  \label{a2}
\end{equation}
and this happens when the value of $\mu $ is approximately $10^{14}$GeV. It
turns out that the value of the remaining coupling constant is not far from
the $\alpha _{\text{QED}}$ and $\alpha _{\text{QCD}}$ so at this energy the
three gauge coupling constants are equal and one can therefore introduce
only one gauge group, simplifying the standard model and reducing the number
of parameters. This hypothesis is called the Grand Unification. One can
pursue this idea further by investigating the behavior of other coupling
constants.

The CKM~\cite{n9,n10} matrix and the quark masses in the standard
model are related to the
couplings of quarks to the Higgs particle. The CKM matrix at the $M_{Z}$
energy does not have a definite symmetry and none of its matrix
elements is vanishing. It has however an interesting hierarchy
property that is best seen in the Wolfenstein parameterization~\cite{W,Buras2} 
\begin{eqnarray}
&&\hat{V}=\left( 
\begin{array}{ccc}
1-\frac{1}{2}\lambda ^{2} & \lambda & A\lambda ^{3}(\rho -i\eta ) \\ 
-\lambda & 1-\frac{1}{2}\lambda ^{2} & A\lambda ^{2} \\ 
A\lambda ^{3}(1-\bar{\rho}-i\bar{\eta}) & -A\lambda ^{2} & 1
\end{array}
\right) , \\
&&\lambda \approx 0.22,\;\;\;\;\bar{\rho}=\rho \left( 1-\frac{\lambda ^{2}}{2%
}\right) ,\;\;\;\bar{\eta}=\eta \left( 1-\frac{\lambda ^{2}}{2}\right) .
\end{eqnarray}
The CKM matrix is obtained from the biunitary diagonalization of the Yukawa
couplings $y_{u}$ and $y_{d}$ of the standard model 
\begin{eqnarray}
&&y_{u}=(U_{u})_{L}^{\dagger }Y^{u}(U_{u})_{R}.\;\;\;\;Y^{u}=\text{diag}
(Y_{u},Y_{c},Y_{t}), \\
&&y_{d}=(U_{d})_{L}^{\dagger }Y^{d}(U_{d})_{R}.\;\;\;\;Y^{d}=\text{diag}
(Y_{s},Y_{s},Y_{b}), \\
&&\hat{V}=(U_{u})_{L}(U_{d})_{L}^{\dagger }.
\end{eqnarray}
The renormalization group flow of the matrices $y_{u}$ and $y_{d}$ of the
Yukawa couplings is determined from the following equations: 
\[
\frac{dy_{u,d}}{dt}=\frac{1}{(4\pi )^{2}}\left\{ {\beta }%
_{u,d}^{(1)}(g_{k}^{2},y_{u}y_{u}^{\dagger },y_{d}y_{d}^{\dagger })+\frac{1}{%
(4\pi )^{2}}{\beta }_{u,d}^{(2)}(g_{k}^{2},y_{u}y_{u}^{\dagger
},y_{d}y_{d}^{\dagger },\lambda _{H})+\cdots \right\} y_{u,d}\,\,\,, 
\]
where $t\equiv \ln (E/\mu )$ is the energy scale parameter. 
Equations for $y_{u}$ and $y_{d}$ have to be
complemented by the following equations for the gauge coupling
constants $g_{i}$, vacuum expectation value $v$ and the Higgs quartic
coupling $\lambda_{H}$ 
\begin{eqnarray*}
\frac{dg_{i}}{dt} &=&\frac{1}{(4\pi )^{2}}\left\{ {\beta }%
_{g_{i}}^{(1)}(g_{i}^{2})+\frac{1}{(4\pi )^{2}}{\beta }%
_{g_{i}}^{(2)}(g_{k}^{2},y_{u}y_{u}^{\dagger },y_{d}y_{d}^{\dagger })+\cdots
\right\} g_{i}, \\
\frac{dv}{dt} &=&\frac{1}{(4\pi )^{2}}\left\{ {\beta }%
_{v}^{(1)}(g_{k}^{2},y_{u}y_{u}^{\dagger },y_{d}y_{d}^{\dagger })+\frac{1}{%
(4\pi )^{2}}{\beta }_{v}^{(2)}(g_{k}^{2},y_{u}y_{u}^{\dagger
},y_{d}y_{d}^{\dagger },\lambda _{H})+...\right\} v, \\
\frac{d\lambda _{H}}{dt} &=&\frac{1}{(4\pi )^{2}}\left\{ {\beta }_{\lambda
_{H}}^{(1)}(g_{k}^{2},y_{u}y_{u}^{\dagger },y_{d}y_{d}^{\dagger })+\frac{1}{%
(4\pi )^{2}}{\beta }_{\lambda _{H}}^{(2)}(g_{k}^{2},y_{u}y_{u}^{\dagger
},y_{d}y_{d}^{\dagger },\lambda _{H})+...\right\} .
\end{eqnarray*}
These equations form a system of coupled nonlinear equations and their
explicit solution is not possible, but one can obtain several interesting
properties of these equations if one makes use of the hierarchy that
is contained in them. The
lowest non trivial order arises when we neglect in the equations the terms
of the order $\lambda ^{4}$ and higher. This approximation is equivalent to
the one loop approximation and the equations have the following 
form~\cite{ref0,refprd}
\begin{equation}
\frac{dg_{i}}{dt}=\frac{1}{(4\pi )^{2}}b_{i}g_{i}^{3},  \label{P1}
\end{equation}
\begin{equation}
\frac{dv_{u,d}}{dt}=\frac{1}{(4\pi )^{2}}[\alpha _{1}^{v_{u,d}}(t)+\alpha
_{3}^{v_{u,d}}
\text{tr}(y_{u}^{\phantom {\ \dagger }}y_{u}^{\dagger })]v_{u,d},
\end{equation}
\begin{eqnarray}
\frac{dy_{u}}{dt} &=&\frac{1}{(4\pi )^{2}}[\alpha _{1}^{u}(t)+\alpha
_{2}^{u}y_{u}^{\phantom {\dagger }}y_{u}^{\dagger }+\alpha _{3}^{u}\text{tr}
(y_{u}^{\phantom {\dagger }}y_{u}^{\dagger })]y_{u}, \\
\frac{dy_{d}}{dt} &=&\frac{1}{(4\pi )^{2}}[\alpha _{1}^{d}(t)+\alpha
_{2}^{d}y_{u}^{\phantom {\dagger }}y_{u}^{\dagger }+\alpha _{3}^{d}\text{tr}
(y_{u}^{\phantom {\dagger }}y_{u}^{\dagger })]y_{d}.
\end{eqnarray}
The various functions and constants in the renormalization group
equations are equal to
\begin{equation}
(b_{1},b_{2},b_{3})=(41/10,-19/6,-7)_{SM},\,\,\,(21/5,-3,-7)_{DHM},\,\,%
\,(33/5,1,-3)_{MSSM},
\end{equation}
\begin{eqnarray}
\alpha _{1}^{u}(t)
&=&-(c_{1}g_{1}^{2}+c_{2}g_{2}^{2}+c_{3}g_{3}^{2}),\,\,\alpha _{2}^{u}=\frac{
3b}{2},\,\,\,\,\alpha _{3}^{u}=3, \\
\alpha _{1}^{d}(t) &=&-(c_{1}^{\prime }g_{1}^{2}+c_{2}^{\prime
}g_{2}^{2}+c_{3}^{\prime }g_{3}^{2}),\,\,\,\alpha _{2}^{d}=\frac{3c}{2}
,\,\,\alpha _{3}^{d}=3a, \\
\alpha _{1}^{v_{u,d}}(t) &=&c_{1(u,d)}^{\prime \prime
}g_{1}^{2}+c_{2(u,d)}^{\prime \prime }g_{2}^{2},\,\,\,\,\,\,\,\,\,\alpha
_{3}^{v_{o}}=-3,\;\alpha _{3}^{v_{u}}=-3,\;\alpha _{3}^{v_{d}}=0,
\end{eqnarray}
with
\[
(a,b,c)=(1,1,-1)_{SM},\;\;(0,1,1/3)_{DHM},\;\;
(0,2,2/3)_{MSSM}
\]
\[
(c_{1},c_{2},c_{3})=(17/20,9/4,8)_{SM},\;\;
(17/20,9/4,8)_{DHM},\;\;
(13/15,3,16/3)_{MSSM},
\]
\[
(c_{1}^{\prime},c_{2}^{\prime },c_{3}^{\prime })=(1/4,9/4,8)_{SM},\;\;
(1/4,9/4,8)_{DHM},\;\;(7/15,3,16/3)_{MSSM},
\]
\[
(c_{1(u,d)}^{\prime \prime },c_{2(u,d)}^{\prime \prime })=
(-9/20,-9/4)_{SM},\;\;
(-9/20,-9/4)_{DHM},\;\;
(-3/20,-3/4)_{MSSM}.
\]
Here(SM) stands for Standard
Model~, (MSSM) for Minimal Supersymmetric Standard Model~ and (DHM) for
Double Higgs Model.

Equations for $g_{i}$ can be solved independently and with these solutions
one can also solve explicitly the equations for $y_{u}$ and $y_{d}$. These
solutions read 
\begin{equation}
g_{i}(t)=g_{i}^{0}\left( 1-\frac{2b_{i}(g_{i}^{0})^{2}(t-t_{0})}{(4\pi )^{2}}%
\right) ^{-1/2},\;\;\;\;g_{i}^{0}\equiv g_{i}(t_{0}),  \label{P2}
\end{equation}
The solutions to the remaining equations are~\cite{ref0,refprd}: 
\begin{equation}
v_{u,d}\left( t\right) =v_{u,d}\left( t_{0}\right) \sqrt{r_{v_{u,d}}^{\prime
\prime }(t)}h_{m}^{\alpha _{3}^{v_{u,d}}}(t),\,\,\,v_{u,d}^{0}=v_{u,d}\left(
t_{0}\right) ,
\end{equation}
\begin{equation}
Y_{u,c}(t)=Y_{u,c}(t_{0})\sqrt{r(t)}h_{m}^{\alpha
_{3}^{u}}(t),\,\,\,Y_{t}(t)=Y_{t}(t_{0})\sqrt{r(t)}\,h_{m}^{\left( \alpha
_{2}^{u}+\alpha _{3}^{u}\right) }(t),
\end{equation}
\begin{equation}
Y_{d,s}(t)=Y_{d,s}(t_{0})\sqrt{r^{\prime }(t)}h_{m}^{\alpha
_{3}^{d}}(t),\,\,\,Y_{b}(t)=Y_{b}(t_{0})\sqrt{r^{\prime }(t)}\,h_{m}^{\left(
\alpha _{2}^{d}+\alpha _{3}^{d}\right) }(t),
\end{equation}
where 
\begin{eqnarray}
r(t) &=&\exp \left[ \frac{2}{(4\pi )^{2}}\int_{t_{0}}^{t}\alpha
_{1}^{u}(\tau )d\tau \right] =\prod_{k=1}^{k=3}\left[ \frac{g_{k}^{2}\left(
t_{0}\right) }{g_{k}^{2}\left( t\right) }\right] ^{\frac{c_{k}}{b_{k}}}, \\
r^{\prime }(t) &=&\exp \left[ \frac{2}{(4\pi )^{2}}\int_{t_{0}}^{t}\alpha
_{1}^{d}(\tau )d\tau \right] =\prod_{k=1}^{k=3}\left[ \frac{g_{k}^{2}\left(
t_{0}\right) }{g_{k}^{2}\left( t\right) }\right] ^{\frac{c_{k}^{\prime }}{%
b_{k}}}, \\
r_{{v_{u,d}}}^{\prime \prime }(t) &=&\exp \left[ \frac{2}{(4\pi )^{2}}%
\int_{t_{0}}^{t}\alpha _{1}^{v_{u,d}}(\tau )d\tau \right]
=\prod_{k=1}^{k=2}\left[ \frac{g_{k}^{2}(t_{0})}{g_{k}^{2}(t)}\right] ^{%
\frac{c_{k(u,d)}^{\prime \prime }}{b_{k}}}, \\
h_{m}(t) &=&\exp \left( \frac{1}{(4\pi )^{2}}\int_{t_{0}}^{t}Y_{t}^{2}(\tau
)d\tau \right) =\left( \frac{1}{1-\frac{2\left( \alpha _{2}^{u}+\alpha
_{3}^{u}\right) }{\left( 4\pi \right) ^{2}}(Y_{t}^{0})^{2}\int_{t_{0}}^{t}r(%
\tau )d\tau }\right)_.^{\frac{1}{2\left( \alpha _{2}^{u}+\alpha
_{3}^{u}\right) }}
\end{eqnarray}

From the solutions for the $y_{u}$ and $y_{d}$ we obtain the
explicit energy dependence of the CKM matrix elements: 
\begin{equation}
|\hat{V}_{td}\left( t\right) |^{2}=\frac{|\hat{V}_{td}^{0}|^{2}}{%
h^{2}+\left( 1-h^{2}\right) |\hat{V}_{td}^{0}|^{2}},\,\,\,\text{where}%
\,\,\,\,\hat{V}_{ij}^{0}\equiv \hat{V}_{ij}\left( t_{0}\right),
\;\;h(t)\equiv\left[ h_{m}\left( t\right)
\right] ^{\alpha _{2}^{d}}\label{II.24a}
\end{equation}
\begin{equation}
|\hat{V}_{cd}\left( t\right) |^{2}=\frac{h^{2}|\hat{V}_{cd}^{0}|^{2}}{%
h^{2}+(1-h^{2})|\hat{V}_{td}^{0}|^{2}},\,\,\,|\hat{V}_{tb}(t)|^{2}=\frac{%
h^{2}|\hat{V}_{tb}^{0}|^{2}}{1+(h^{2}-1)|\hat{V}_{tb}^{0}|^{2}},
\label{II.24b}
\end{equation}
\begin{equation}
|\hat{V}_{ub}\left( t\right) |^{2}=\frac{|\hat{V}_{ub}^{0}|^{2}}{1+(h^{2}-1)|%
\hat{V}_{tb}^{0}|^{2}}.  \label{II.24d}
\end{equation}
All the remaining CKM matrix elements can be obtained from the unitarity.
From these equations the following transformation laws of the Wolfenstein
parameters are obtained: 
\begin{equation}
A\left( t\right) =\frac{A}{h(t)},\,\,\,\,\,\,\,\,\,\,\text{and\thinspace
\thinspace \thinspace \thinspace \thinspace \thinspace \thinspace }\lambda
,\rho ,\eta \text{ are invariant.}  \label{69}
\end{equation}
The evolution of the CP violation Jarlskog invariant~\cite{n12a} is
also very simple
\begin{equation}
J=\Im \left[ \hat{V}_{ud}\hat{V}_{cs}\hat{V}_{us}^{*}\hat{V}_{cd}^{*}\right]
=\Im \left[ \hat{V}_{ud}\hat{V}_{tb}\hat{V}_{ub}^{*}\hat{V}_{td}^{*}\right] ,
\end{equation}
\begin{equation}
J=\frac{J_{0}}{\left| 1+(h^{2}-1)|\hat{V}_{tb}^{0}|^{2}\right| }\approx 
\frac{J_{0}}{h^{2}}.  \label{66}
\end{equation}
All these results can be summarized by the following simple evolution of the
CKM matrix 
\begin{equation}
\left( 
\begin{array}{ccc}
\displaystyle\hat{V}_{ud}^{0} & \displaystyle\hat{V}_{ud}^{0} & \displaystyle%
\hat{V}_{ub}^{0} \\ 
\displaystyle\hat{V}_{cd}^{0} & \displaystyle\hat{V}_{cs}^{0} & \displaystyle%
\hat{V}_{cb}^{0} \\ 
\displaystyle\hat{V}_{td}^{0} & \displaystyle\hat{V}_{ts}^{0} & \displaystyle%
\hat{V}_{tb}^{0}
\end{array}
\right) \longrightarrow \left( 
\begin{array}{ccc}
\displaystyle\hat{V}_{ud}^{0} & \displaystyle\hat{V}_{ud}^{0} & \displaystyle%
\frac{\hat{V}_{ub}^{0}}{h} \\ 
\displaystyle\hat{V}_{cd}^{0} & \displaystyle\hat{V}_{cs}^{0} & \displaystyle%
\frac{\hat{V}_{cb}^{0}}{h} \\ 
\displaystyle\frac{\hat{V}_{td}^{0}}{h} & \displaystyle\frac{\hat{V}_{ts}^{0}%
}{h} & \displaystyle\hat{V}_{tb}^{0}
\end{array}
\right)  \label{evCKM}
\end{equation}

We finally summarize our conclusions formulating the following theorems:\\[3pt]
\textit{Theorem 1:} The one loop renormalization group flow of the CKM
matrix depends on only one universal function of the energy $h(t)$.\\[1pt]
\textit{Theorem 2:} The ratios of the down quark masses $m_{d}/m_{b}$ and $%
m_{s}/m_{b}$ are the functions of only $h(t)$.\\[1pt]
\textit{Theorem 3:} The ratio of the up and down quark masses $m_{u}/m_{c}$
and $m_{u}/m_{c}$ are energy independent.\\[1pt]
\textit{Theorem 4:} The next order non vanishing corrections for the flow of
the CKM matrix are of the order $\lambda ^{5}$.\\[1pt]
\textit{Theorem 5:} The unitarity triangle is invariant trough the evolution.
\begin{theacknowledgments}
We gratefully acknowledge the financial support from
CONACYT-Proyecto-ICM (Me\-xi\-co). S.R.J.W. also thanks to ``Comisi\'{o}n
de Operaci\'{o}n y Fomento de Actividades Acad\'{e}micas'' (COFAA) from
Instituto Polit\'{e}cnico Nacional.
\end{theacknowledgments}

\end{document}